\documentclass[11pt,,a4paper]{article} 
\pdfoutput=1
\usepackage{jcappub}
\usepackage{graphics} 
\usepackage{epsf}  
\usepackage{eucal} 
\usepackage{amssymb}
\usepackage{amsmath} 
\usepackage{graphicx} 
\usepackage{bm}
\usepackage{dcolumn} 
\usepackage{epsfig} 
\usepackage{hyperref}
\usepackage{flafter} 
\usepackage{xcolor}

\newcommand{\be}{\begin{equation}}
\newcommand{\ee}{\end{equation}}
\newcommand{\bea}{\begin{eqnarray}}
\newcommand{\eea}{\end{eqnarray}}

\begin{document}

\title{Observational implications of mattergenesis during inflation}

\author[a]{Mar Bastero-Gil} \emailAdd{mbg@ugr.es} \affiliation[a]{Departamento
  de F\'{\i}sica Te\'orica y del Cosmos, Universidad de Granada,
  Granada-18071, Spain}

\author[b]{Arjun Berera} \emailAdd{ab@ph.ed.ac.uk} \affiliation[b]{SUPA, School
  of Physics and Astronomy, University of Edinburgh, Edinburgh, EH9
  3JZ, United Kingdom}

\author[c]{Rudnei O. Ramos} \emailAdd{rudnei@uerj.br}
\affiliation[c]{Departamento de F\'{\i}sica Te\'orica, Universidade do
  Estado do Rio de Janeiro, 20550-013 Rio de Janeiro, RJ, Brazil}

\author[d]{Jo\~ao G. Rosa}
\emailAdd{joao.rosa@ua.pt} 
\affiliation[d]{Departamento de F\'{\i}sica da Universidade de Aveiro and I3N, Campus de Santiago, 3810-183 Aveiro, Portugal} 

\date{\today}

\arxivnumber{1404.4976}

\abstract{
The observed baryon asymmetry, as well as potentially an asymmetry in the dark matter sector, can be produced through dissipative particle production during inflation. A distinctive feature of this mechanism is the generation of matter isocurvature perturbations that are fully (anti-)correlated with the dominant adiabatic curvature perturbations. We show that chaotic warm inflation models yield anti-correlated isocurvature modes that may partially or even completely screen the contribution of primordial gravity waves to the CMB temperature power spectrum. The tensor-to-scalar ratio inferred from the latter may thus be parametrically smaller than the one deduced from B-mode polarization maps, which is particularly relevant in the light of the recently announced results of the BICEP2 experiment.
}


\maketitle
\section{Introduction}

The BICEP2 experiment \cite{Ade:2014xna} has recently reported evidence for a large tensor-to-scalar ratio $r=0.16^{+0.06}_{-0.05}$ (or $r=0.20^{+0.07}_{-0.05}$ without foreground dust subtraction) from the observation of B-mode polarization in the cosmic microwave background (CMB) at degree angular scales. While this is good news for the inflationary paradigm \cite{Guth, Albrecht, Linde}, which predicts a primordial tensor component in the CMB spectrum,  BICEP2's value seems to be in tension with the constraint on the tensor-to-scalar ratio reported by the Planck collaboration last year  \cite{Ade:2013uln}. The Planck collaboration has, in particular, placed an upper bound $r < 0.11$ (95\% CL), assuming that primordial scalar curvature perturbations are described solely by an adiabatic component with a simple power-law spectrum, i.e.~no running of the spectral index. 

On the one hand, Planck has also confirmed a significant deficit of power on large angular scales with respect to their best-fit $\Lambda$CDM model, with a primordial spectrum characterized by a constant red-tilted spectral index, so that any additional contributions like gravity waves are naturally rather constrained. On the other hand, any modification of the primordial spectrum that tends to reduce the power on large scales will help relaxing the above constraint on $r$.  Several possibilities were already mentioned by the Planck collaboration, and they have been further explored in view of the BICEP2 result, for example a negative running of the scalar spectral index \cite{Czerny:2014qqa,McDonald:2014kia}; sterile neutrinos as extra relativistic degrees of freedom \cite{Giusarma:2014zza, Zhang:2014dxk,Dvorkin:2014lea}; a blue-tilted tensor spectrum \cite{bluetensor1, bluetensor2}; or isocurvature perturbations \cite{Kawasaki:2014lqa,Kawasaki:2014fwa}. In particular the tension in the bound on the tensor-to-scalar ratio $r$ between Planck and BICEP2 can be resolved by introducing isocurvature perturbations that are anti-correlated with the main adiabatic component.

When the cosmological baryon asymmetry is produced through dissipative particle production during inflation, a mechanism known as {\it warm baryogenesis} \cite{BasteroGil:2011cx}, super-horizon fluctuations of the inflaton field become imprinted in the resulting baryon-to-entropy ratio. This then generates baryon isocurvature fluctuations that are either fully correlated or anti-correlated with the main adiabatic curvature perturbations generated during inflation (whereas for other post-inflationary baryogenesis scenarios these will be uncorrelated, see e.g.~\cite{otherbaryo1, otherbaryo2, otherbaryo3}). However, in deriving the bounds on the tensor-to-scalar ratio $r$ from CMB temperature anisotropies, the Planck collaboration has assumed that primordial scalar curvature perturbations are described uniquely by an adiabatic component. The effects of any other component such as baryon isocurvature modes are then necessarily absorbed into an effective tensor-to-scalar ratio, $r_{eff}$, which is smaller than the true tensor contribution if the additional components are anti-correlated with the dominant adiabatic modes.

 This effective screening goes along the lines proposed in \cite{Kawasaki:2014lqa}, where it was shown that anti-correlated cold dark matter (CDM) isocurvature perturbations could completely cancel the tensor contribution to the temperature power spectrum. In this work, we show that in chaotic models of warm baryogenesis an either partial or even full screening is naturally present and may reconcile the BICEP2 detection of B-mode polarization with the upper bound on the tensor-to-scalar ratio placed by Planck.  A partial screening would, in particular, be interesting if there is future evidence for a non-zero tensor-to-scalar ratio in the temperature power spectrum that is somewhat smaller than the value inferred from the polarization data. This screening can, in fact, be effective for a wide range of values for the tensor-to-scalar ratio and is not inherent to the large value obtained by the BICEP2 collaboration, which is presently under scrutiny.

CDM isocurvature modes have been considered in several contexts, such as axion or curvaton models \cite{Lyth:2002my,
  Byrnes:2014xua, Ashoorioon:2009wa, Ashoorioon:2014jja}. Being more suppressed, the baryonic contribution has received relatively less attention in the literature, but since the corresponding isocurvature perturbation is $S_b=\delta \eta_s/\eta_s$, where $\eta_s$ is the baryon-to-entropy ratio, their presence is necessarily connected to the mechanism generating a cosmological baryon asymmetry. Hence, identifying such a contribution in the CMB power spectrum may provide crucial information about baryogenesis. Dissipation may also produce an asymmetry in the CDM sector, provided that CDM particles are characterized by a global charge that is not conserved in the relevant dissipative processes. In this work, we thus consider the more general {\it warm mattergenesis} scenario, where isocurvature modes can be generated in both the baryon and CDM fluids. We show, in particular, that the spectrum of these perturbations is entirely determined by the underlying inflationary model and particle interactions, yielding concrete predictions that may be further tested in the near future.
  
The tension between BICEP2 and Planck naturally leads one to question whether the simplest inflationary models can really describe the observable universe, with only a single dynamical scalar field driving accelerated expansion in an almost perfect vacuum. A particular assumption underlying these simple models is that interactions between the inflaton field and other degrees of freedom have a negligible effect during the accelerating phase, only becoming relevant after the inflaton field has exited the slow-roll regime and the Hubble rate has decreased sufficiently. However, this is not necessarily the case in general, and inflation may in fact occur in a dissipative regime, as is actually the case for most dynamical systems in Nature. 

Dissipation is a natural consequence of interactions between the inflaton and other fields and may have a plethora of interesting effects. By dissipating its energy, the inflaton will roll more slowly down its potential, alleviating the need for very flat potentials. At the same time, the dissipated energy will necessarily source other fluids, counteracting the diluting effect of the quasi-exponential expansion and thus sustaining a non-vacuum state during inflation. Although this may be an arbitrary state, most discussions in the literature consider the case where the particles produced by dissipation thermalize within a Hubble time, leading to a near-equilibrium configuration characterized by a slowly-evolving temperature $T$. This idea is thus generically known as {\it warm inflation} \cite{Berera:1995wh, Berera:1995ie, Berera:2008ar}, as opposed to the supercooled regime of the conventional non-dissipative models.

Warm inflation has several interesting features besides slowing down the inflaton's motion. Although sub-leading during the slow-roll phase, the relative abundance of the radiation bath may slowly increase, eventually taking over the inflaton's vacuum energy at the end of inflation. This yields a smooth exit into the standard Big Bang cosmology, with no need for a separate reheating period since the universe never actually supercooled. It is also a well-known result that dissipation induces fluctuations in a system, the inflaton being no exception, and as a result the presence of dissipative effects necessarily changes the spectrum of primordial density perturbations. By looking for signatures of dissipative effects in the CMB, one may thus hope to learn something about the interactions between the inflaton and other degrees of freedom, for which there is little hope with a separate reheating period occurring 50-60 e-folds after the relevant CMB scales left the Hubble horizon during inflation. A finite temperature during inflation has also been recently argued  as a means to stabilize the Higgs field during inflation \cite{Espinosa:2007qp, Fairbairn:2014zia}, in particular since the high inflationary scale suggested by the BICEP2 results would render the electroweak vacuum unstable. 

The combination of thermal and dissipative effects may also lead to several other processes during inflation, including the production of baryons and dark matter as originally proposed in  \cite{BasteroGil:2011cx}, which is the main focus of this work. Dissipation naturally produces particles in an out-of-equilibrium fashion, so that if baryon number (or an analogous dark matter number), as well as $C$ and $CP$, are not conserved in the interactions between the inflaton and other fields, all the conditions established by Sakharov \cite{Sakharov:1967dj} for the asymmetric production of baryons and anti-baryons are satisfied. We discuss below how this mechanism is concretely realized within the best understood quantum field theory framework for warm inflation, where one of the most significant consequences is the fact that the resulting baryon/CDM asymmetry depends on the inflaton field value. Hence, the very same inflaton fluctuations that generate adiabatic curvature perturbations will also generate super-horizon perturbations in the baryon/CDM-to-entropy ratio, i.e.~isocurvature modes as anticipated above.  

In this work we focus on the relation between matter isocurvature modes and the resulting effective tensor-to-scalar ratio in models which can produce a large value of $r$, i.e.~chaotic models with a power-law potential. In \cite{Bartrum:2013fia} it was shown that a quartic chaotic potential can be reconciled with Planck when dissipative effects are taken into account, and the interactions can maintain a thermal distribution of inflaton particles. We will now show that, if the observed baryon asymmetry is produced during inflation, observational predictions can be consistent with both BICEP2 and Planck, in the sense of having a detection of tensors on large scales screened by the isocurvature contribution in the analyses realized by Planck so far. In fact, for values of $r$ somewhat below the current BICEP2 range, which for warm chaotic models correspond to larger temperatures, this screening can render gravity waves unobservable in the CMB temperature power spectrum. We will show, moreover, that tensor screening can be effective for matter isocurvature perturbations within the bounds placed by Planck \cite{Ade:2013uln}.

This article is organized as follows. In section II we revise the basic mechanism and dynamics of warm inflation and the associated predictions for the primordial perturbation spectrum. We study both the quartic and quadratic warm inflation models and compare their observational predictions. In section III we explore the effect of isocurvature modes in screening the tensor-to-scalar ratio for these models. We summarize and discuss our main conclusions in section IV.  


\section{Warm inflation dynamics and primordial perturbation spectrum}

Warm inflation is based on the inclusion of dissipative processes in the dynamics of inflation as mentioned in the Introduction section. While this may in general result in different non-equilibrium configurations, the simplest case corresponds to a scenario where the particles produced by dissipation thermalize faster than Hubble expansion, leading to a quasi-adiabatic and near-equilibrium evolution. The leading effect results in an effective friction term $\Upsilon$ in the inflaton's equation of motion \cite{Moss:1985wn, Yokoyama:1987an, Berera:1995wh, Berera:1995ie, Berera:2008ar}, which modifies the evolution of both the classical homogeneous background and the associated fluctuations.

If the produced particles are relativistic, i.e.~light compared to the ambient temperature, the inflaton's energy density is dissipated into a nearly-thermal radiation bath, so that the background evolution is described by the system of equations:
\bea \label{background_eqs}
\ddot\phi+3H\dot\phi+\Upsilon\dot\phi+V'(\phi) &=&0~,\label{inflaton_eq}\\
\dot\rho_R+4H\rho_R&=&\Upsilon\dot\phi^2~, \label{radiation_eq} 
\eea
with $\rho_R=(\pi^2/30)g_*T^4$ for $g_*$ relativistic degrees of
freedom. In the overdamped or slow-roll regime, these reduce to:
\bea
3H(1+Q)\dot\phi &\simeq& -V'(\phi) \,,\\
4H\rho_R &\simeq&\Upsilon\dot\phi^2 \,,
\eea
where $Q=\Upsilon/3H$. This regime can be sustained if the generalized conditions $\epsilon,|\eta|\ll1+Q$ on the slow-roll
parameters $\epsilon=(M_P^2/2)(V'/V)^2$ and $\eta=M_P^2 V''/V$ can be satisfied. For strong dissipation, $Q\gtrsim 1$, these conditions allow for inflationary expansion with $\epsilon, |\eta|\gtrsim 1$, which relaxes the requirement of finely-tuned scalar potentials \cite{Berera:1995ie}. 

In the slow-roll regime, we obtain $\rho_R/\rho_\phi\simeq (\epsilon/2)Q/(1+Q)^2\ll1$. Radiation is then a sub-dominant
component, which is necessary for accelerated expansion, although the temperature of the thermal bath may exceed the Hubble rate, $T>H$, even for weak dissipation, $Q\ll1$. However, at the end of inflation we have $\epsilon\sim 1+Q$ and so $\rho_R/\rho_\phi\sim
(1/2)Q/(1+Q)$. This shows that if dissipative effects become larger as inflation proceeds and $Q\gtrsim 1$ at the end, the relative abundance of radiation will naturally increase and eventually become the dominant component, leading to a smooth exit into the standard cosmological evolution.

The dissipation coefficient may in general be computed in an adiabatic approximation from the quantum self-energy of the inflaton field. It is typically difficult for the inflaton to dissipate its energy directly into light degrees of freedom, on the one hand because this would generate large thermal corrections to the inflaton's mass and, on the other hand, since fields coupled to the inflaton typically acquire a very large mass unless the associated couplings are sufficiently suppressed \cite{highT, YL}. A radiation bath may nevertheless be produced if the inflaton dissipates its energy into heavy fields that subsequently decay into light degrees of freedom, which in supersymmetric (SUSY) theories is realized via a renormalizable superpotential of the form  \cite{Berera:2002sp, Moss:2006gt, BasteroGil:2010pb, BasteroGil:2012cm}: 
\begin{equation} \label{superpotential}
W=f(\Phi)+{g\over2}\Phi X^2+ {h\over 2}XY^2~,
\end{equation}
where the inflaton field corresponds to the scalar component of the chiral multiplet $\Phi$, $\phi=\sqrt{2}\langle\Phi\rangle$, and the associated  scalar potential is $V(\phi)=|f'(\phi)|^2>0$, which leads to SUSY breaking during inflation. On the one hand, as anticipated above, the bosonic and fermionic components of the $X$ fields gain a mass $m_X\simeq(g/\sqrt{2})\phi$ during inflation, which is typically large, in particular for large-field models. On the other hand, it is not difficult to see that the $Y$ bosons and fermions remain massless at tree-level and that the last Yukawa term in the superpotential allows for the decays $\chi\rightarrow yy, \psi_y\psi_y$ and $\psi_\chi\rightarrow y\psi_y $, with $(\chi, \psi_\chi)$ and $(y,\psi_y)$ denoting the scalar and fermionic fields within the $X$ and $Y$ chiral multiplets.

The computation of the dissipation coefficient can be performed using standard techniques in thermal field theory and has been done in the literature \cite{Moss:2006gt, BasteroGil:2010pb, BasteroGil:2012cm}. The underlying physical mechanism is not difficult to understand. Interactions with the $X$ fields modify the inflaton's effective action at the quantum level, the leading 1-loop effect being the creation and annihilation of $X$-pairs coupled to the background condensate. While for a static background field this only leads to radiative corrections to the effective scalar potential, in the dynamical case this induces time non-local corrections to the full effective action, with creation and subsequent annihilation of $X$ pairs occurring at different times, at which the background field takes different values. For the quasi-adiabatic evolution typical of the slow-roll regime, the field changes by $\Delta\phi\simeq \dot\phi \Delta t$, leading to the friction coefficient $\Upsilon\dot\phi$ in Eq.~(\ref{background_eqs}). The dynamical nature of the field then yields an asymmetry between creation and annihilation processes and hence a net particle production, which sources the radiation bath since the $X$ particles are unstable against decay into the $Y$ sector.

The dissipation coefficient receives contributions from both on-shell and virtual $\chi$ and $\psi_\chi$ modes. As shown in \cite{BasteroGil:2010pb, BasteroGil:2012cm, Moss:2006gt}, the leading contribution for heavy fields $m_X\gtrsim T$ is given by off-shell scalar $\chi$ modes that dominantly decay into $y$ scalars, yielding a dissipation coefficient:
\begin{equation} \label{dissipation_coefficient}
\Upsilon=C_\phi {T^3\over \phi^2}~,\qquad C_\phi\simeq  {1\over4}\alpha_h N_X~,
\end{equation}
for $\alpha_h=h^2N_Y/4\pi\lesssim1$ and $N_{X,Y}$ chiral multiplets.  Fermionic interactions yield different, subdominant contributions to the
dissipation coefficient \cite{BasteroGil:2010pb}.  We note that the dissipation coefficient depends on both the field and temperature, which is in fact generically the case, making it a dynamical quantity that evolves during inflation. One may then e.g.~consider scenarios where dissipation has negligible effects 50-60 e-folds before the end of inflation, $Q_*\ll 1$, but becomes stronger as inflation proceeds, eventually leading to the smooth exit into a radiation-dominated universe as described above. Also, we note that the strength of the dissipation coefficient depends on the field multiplicities in both the $X$ and $Y$ sectors but only on the Yukawa coupling $h$. The independence of the constant parameter $C_\phi$ on the coupling $g$ is particularly important, since this coupling controls the strength of radiative corrections to the inflaton potential, which are partially cancelled by the underlying SUSY and have the form:
\begin{equation} \label{1-loop}
{ \Delta V^{(1)}\over V}\simeq {\alpha_g\over8\pi}\log(m_X^2/\mu^2)~,
\end{equation}
where $\alpha_g=g^2N_X/4\pi$ and $\mu$ is the renormalization scale. This means that one may consider perturbative models where $\alpha_{g,h}\lesssim 1$, which are therefore fully controllable and renormalizable, while increasing dissipative effects for $N_X\gg 1$ and $g\ll 1$. In particular, from the slow-roll equations one can easily deduce that:
\begin{equation} \label{T/H}
{T\over H}={C_\phi\over 4C_R}\left({d\log\phi\over dN_e}\right)^2 \simeq {15\over8\pi^2}\left({\Delta\phi\over \phi}\right)^2 N_e^{-2} {N_X\over g_*}~,
\end{equation}
so that to keep $T>H$, i.e.~above the Hawking temperature of the de Sitter horizon, at least for the last 50-60 e-folds of inflation one typically requires $N_X\sim10^4-10^6$ fields as explicitly obtained in several specific models \cite{BasteroGil:2009ec, Cerezo:2012ub}. Although no analysis of dissipative effects for $T<H$ has been performed to date, it is clear that smaller field multiplicities would be required to achieve $T>H$ only during the last few e-folds of inflation. Nevertheless, as we discuss below many of the most significant observational effects arise when this condition is satisfied. 

While these numbers may seem somewhat large compared to the number of known particle species, they are not at all implausible given that the inflaton is typically taken as a singlet field under the Standard Model gauge group or its extensions, being {\it a priori} allowed to interact democratically with the plethora of additional fields typically found in proposed high-energy completions of the Standard Model (SUSY, extra dimensions, GUT, string/M-theory, etc). Moreover, the BICEP2 result points towards an inflationary scale close to the GUT scale, i.e.~about 13 orders of magnitude above the current LHC energy reach, so it is not really difficult (and arguably desirable) to envisage large numbers of new particle species arising in between. As an example, concrete realizations of Eq.~(\ref{superpotential}) are naturally found in brane-antibrane systems, where large $N_X\sim N_c^2$ can be obtained from a moderately large number $N_c$ of D-branes \cite{BasteroGil:2011mr}. Since most of these may annihilate at the end of inflation as a result of a tachyonic instability, much like in hybrid inflation models, only a few amongst the heavy fields mediating dissipative effects during inflation will have an effect in the post-inflationary universe, which may also be the case in other realizations of warm inflation. The inclusion of on-shell mode contributions to the dissipative coefficient may nevertheless reduce the required field multiplicities \cite{BasteroGil:2012cm, Bastero-Gil:2013owa}.

Given the functional  form of the dissipative coefficient, one can derive the slow-roll relation:
\be
Q^{1/3}  (1 + Q )^2 = 2 \epsilon_\phi \left(\frac{C_\phi}{3}\right)^{1/3} 
\left(\frac{30 C_\phi}{4 \pi^2 g_*}\right)
\left(\frac{m_P}{\phi}\right)^{8/3} \left(\frac{H}{m_P}\right)^{2/3}~.
\, \label{qphi}
\ee
In particular, for monomial potentials $V\propto \phi^p$, $H\propto \phi^{p/2}$, for which the inflaton decreases during
inflation, dissipation increases over the Hubble rate for $p \leq 14$. This increase will be larger the smaller the power in the potential. Moreover,
for both the quartic ($p=4$) and the quadratic ($p=2$) potentials, inflation ends with $Q>1$ even taking the minimum value of $Q_*$ at horizon-crossing consistent with $T_*>H_*$.  Overall, this extra friction makes inflation last longer no matter what the value of $Q_*$ is.   

The heavy $X$ fields may also decay into inflaton particle states through $\chi\rightarrow yy\phi$, which is, however, a sub-dominant process when computing the dissipative coefficient \cite{BasteroGil:2012cm}.  Nevertheless, dissipative processes may maintain a non-trivial distribution of inflaton particles during inflation, and for sufficiently fast interactions this should approach the equilibrium Bose-Einstein distribution $n_{BE} (k, T) = (e^{k/aT}- 1)^{-1}$ at the same temperature $T$ as the remaining light degrees of freedom. The efficiency of inflaton thermalization processes can be estimated by comparing the relevant interaction rates with the Hubble parameter. This was done in \cite{Bartrum:2013fia} for the quartic chaotic model, where it was found that inflaton particles could be produced and thermalize sufficiently fast when the effective couplings $\alpha_{g,h}$ are not too small. This is also the case for other forms of the potential since in general one finds:
\be
\frac{\Gamma_\phi}{H} \simeq \frac{\alpha_g\alpha_h}{64\pi} \frac{m_X}{T}\frac{T}{H}\,,
\ee
with $T/H,\, m_\chi/T \gtrsim 1$. This implies that inflaton fluctuations are not necessarily in an underlying vacuum state in warm inflation and that the computation of the spectrum of primordial density perturbations should account for non-trivial occupation numbers.
 
As mentioned above, the presence of dissipative effects necessarily affects the evolution of fluctuations, and the fluctuation-dissipation (FD) relation implies that inflaton perturbations are sourced by a gaussian white noise term $\xi_k$ \cite{Berera:1995wh, Berera:1995ie, Berera:1999ws, Hall:2003zp}: 
\begin{equation} \label{fluctuation_eq}
\delta\ddot\phi_k+3H(1+Q)\delta\dot\phi_k+{k^2\over
  a^2}\delta\phi_k\simeq \sqrt{2\Upsilon T}a^{-3/2}\xi_k~,
\end{equation}
in the slow-roll regime. Taking into account that the inflaton may have a  generic  phase-space distribution $n_*$ at the time when observable CMB scales leave the horizon during inflation, one then obtains the dimensionless power spectrum \cite{Berera:1995wh, Berera:1995ie, Moss:1985wn, Berera:1999ws, Hall:2003zp, Ramos:2013nsa}: 
\be \label{scalar_spectrum}
\Delta_\mathcal{R}^2=
\left({H_*\over\dot\phi_*}\right)^2\left({H_*\over 2\pi}\right)^2
\left(1+2n_*+{2\sqrt3\pi Q_*\over\sqrt{3+4\pi Q_*}}{T_*\over H_*}\right)~,
\ee
where all quantities are evaluated at horizon crossing, and $n_* = n_{BE}(a_* H_*, T_*)$ when inflaton particles are produced and thermalized fast enough. This expression yields the standard supercooled inflation result in the limit $n_*, Q_*, T_*\rightarrow 0$. On the other hand, it is only valid in the weak dissipative regime, with $Q_* \lesssim 0.01$, such that the coupling between inflaton and radiation fluctuations can be neglected. Otherwise, the coupled system of fluctuations leads to a ``growing mode'' in the spectrum, with  the amplitude of inflaton and
radiation fluctuations being amplified before freeze-out, so that the spectrum is enhanced by a factor $\Delta_{\mathcal R} \propto Q^\alpha$, with
$\alpha>1$ \cite{Graham:2009bf,BasteroGil:2011xd}. The primordial curvature amplitude can always be adjusted to the COBE normalization observational value by a proper choice of the inflaton self-coupling and/or mass. However, this effect is more pronounced on small scales if $Q$ grows during inflation as for the monomial potentials above \cite{Bastero-Gil:2014jsa}. This would then lead to a too blue-tilted spectrum, which seems to be ruled out by observations. We will therefore concentrate in the following on the regime $Q_* \lesssim 0.01$.  

Taking $Q_*\ll 1$, the scalar power spectrum Eq.~(\ref{scalar_spectrum}) becomes:
\begin{equation} \label{scalar_power_spectrum_full}
\Delta_\mathcal{R}^2\simeq\left({H_*\over\dot\phi_*}\right)^2\left({H_*\over 2\pi}\right)^2\left[1+2n_*+2\pi Q_*{T_*\over H_*}\right]~,
\end{equation}
where the second term within brackets gives the effect of non-trivial inflaton occupation numbers and the last term is the leading effect of FD dynamics. Both corrections are positive, making the amplitude of curvature perturbations larger in warm inflation than in the corresponding supercooled scenarios for the same field and parameter values. Due to their weak coupling to matter, we do not expect the spectrum of primordial gravity waves to be modified by FD, nor will gravitons be produced with non-trivial occupation numbers. The tensor spectrum therefore has the standard vacuum form:
\be \label{tensor_power_spectrum}
\Delta_t^2=\frac{2 H_*^2}{\pi^2 M_P^2}~,
\ee
and as a result the tensor-to-scalar ratio is reduced. In particular, the relation between $r$ and the tensor tilt $n_t=-2\epsilon_*$ is no longer given by the standard supercooled expression $r=8|n_t|$, yielding instead:  
\begin{equation} \label{consistency}
r\simeq{8|n_t|\over 1+2n_*+2\pi Q_*T_*/H_*}~,
\end{equation}
where $n_t=-2\epsilon_*$ is the tensor index. This gives a modified consistency relation for warm inflation and which one may hope to use to distinguish it from supercooled regimes in a way that is actually independent of the form of the dissipation coefficient and hence of the underlying particle physics model. Measuring any deviations from $r=8|n_t|$ may then suggest the presence of dissipative effects, and thus interactions between the inflaton and other fields, which may be complemented by measuring other observables. We note that even if $Q_*\ll 1$ and inflaton occupation numbers are negligible, this modification may nevertheless be significant for $T_*>H_*$, as in the quartic chaotic model studied in \cite{Bartrum:2013fia} that we revisit in this work in the context of warm baryogenesis/mattergenesis.

We may then distinguish two limiting cases. Firstly, if inflaton particles are not produced sufficiently fast, any initial non-trivial occupation numbers will be quickly redshifted away by Hubble expansion, even though the radiation bath of $Y$ particles remains in a near-equilibrium configuration. We may then neglect the effect of $n_*$ on the spectrum and inflaton fluctuations are in this case mainly driven by the near-equilibrium FD term in Eq.~(\ref{scalar_spectrum}). This yields a spectral index for $Q_*\ll1$:
\be  \label{scalar_index_vac} 
n_s-1 \simeq  
2\eta_{*}-6\epsilon_{*}+\frac{2\kappa_*}{1+\kappa_*}\left(7\epsilon_{*}-4\eta_{*}+5\sigma_{*}\right)~,
\end{equation}
where $\kappa\equiv2\pi QT/H$ and $\sigma=M_P^2 V'/(\phi V)$. The corresponding running of the spectral index is given by:
\bea  \label{running_vac}
n_s'&\simeq & -24\epsilon_*^2+16\epsilon_*\eta_*-2\xi_*^2+{\kappa_*\over(1+\kappa_*)^2}\left(14\epsilon_*-8\eta_*+10\sigma_*\right)^2\nonumber\\
&+&{2\kappa_*\over1+\kappa_*}\left(28\epsilon_*^2-17\epsilon_*\eta_*+5\sigma_*^2+10\sigma_*\epsilon_*-5\sigma_*\eta_*+4\xi_*^2\right)~, 
\eea
where $\xi^2=M_P^4 V'(\phi) V'''(\phi)/ V(\phi)^2$. 

In the opposite limit where inflaton fluctuations remain in equilibrium with the $Y$ sector fields, with $n_*\simeq n_{BE*}(a_*H_*, T_*)$, we obtain to lowest order in $Q_*$:
\begin{equation} \label{scalar_index_thermal}
n_s-1\simeq 2\sigma_{*}-2\epsilon_{*} + 2 \left[1-\frac{H_*/T_*}{\sinh
  (H_*/T*)}\right] (\eta_* + 2 \epsilon_*- \sigma_*)~.
\end{equation}
Note that the spectral index does not effectively depend on the curvature of the potential when $T_*/H_* \gtrsim1$, which signals a significant departure from the supercooled inflation predictions. The $\eta$ parameter only affects the running of the spectral index, which is in this case given by:

\begin{equation} \label{running_thermal}
n_s'\simeq 2\sigma_{*}(\sigma_{*}+2\epsilon_{*}-\eta_{*})-4\epsilon_{*}(2\epsilon_{*}-\eta_{*})~.
\end{equation}

\begin{figure}[t] 
\begin{tabular}{ccc}
\centering\includegraphics[scale=0.26]{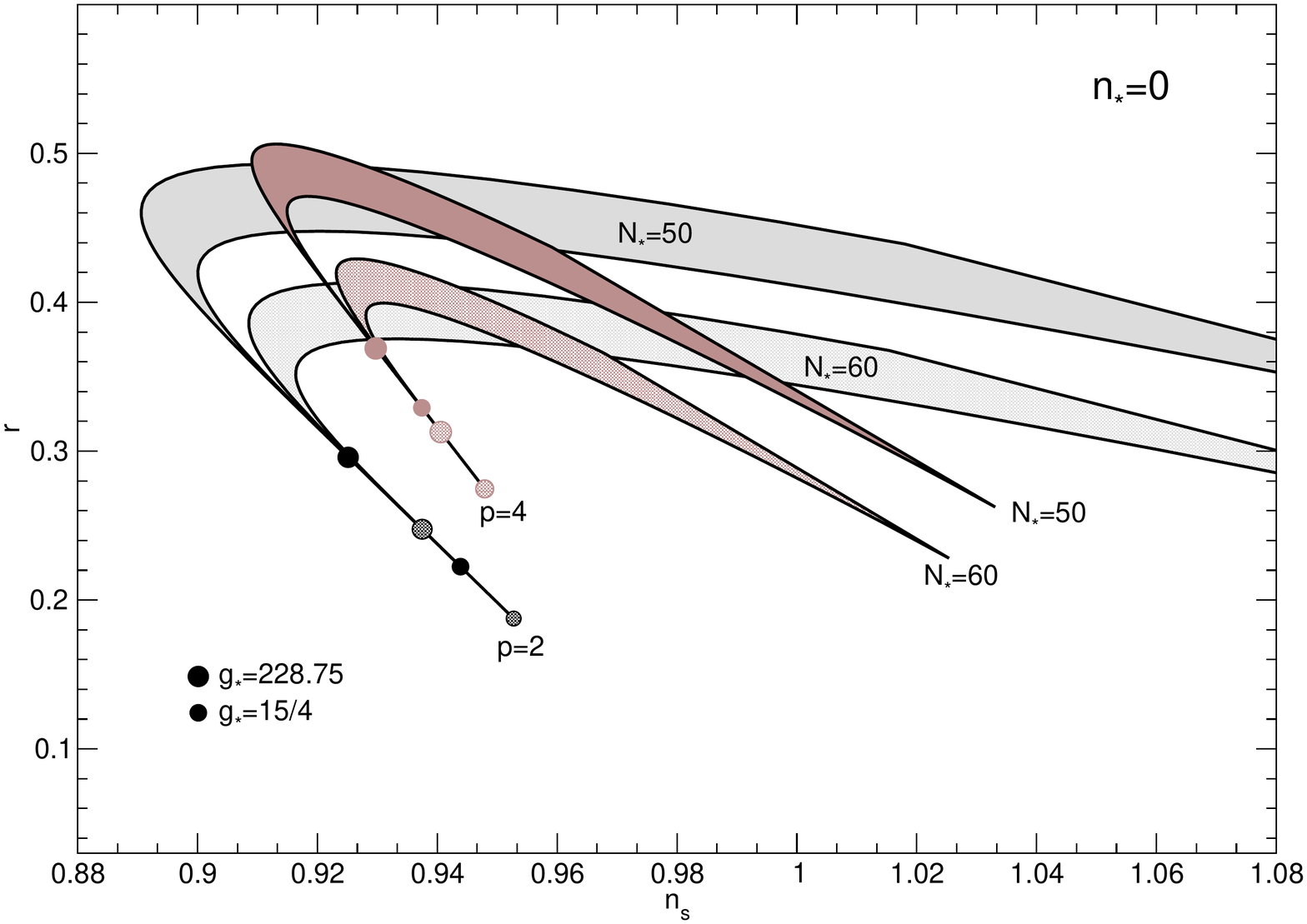} &&
\centering\includegraphics[scale=0.26]{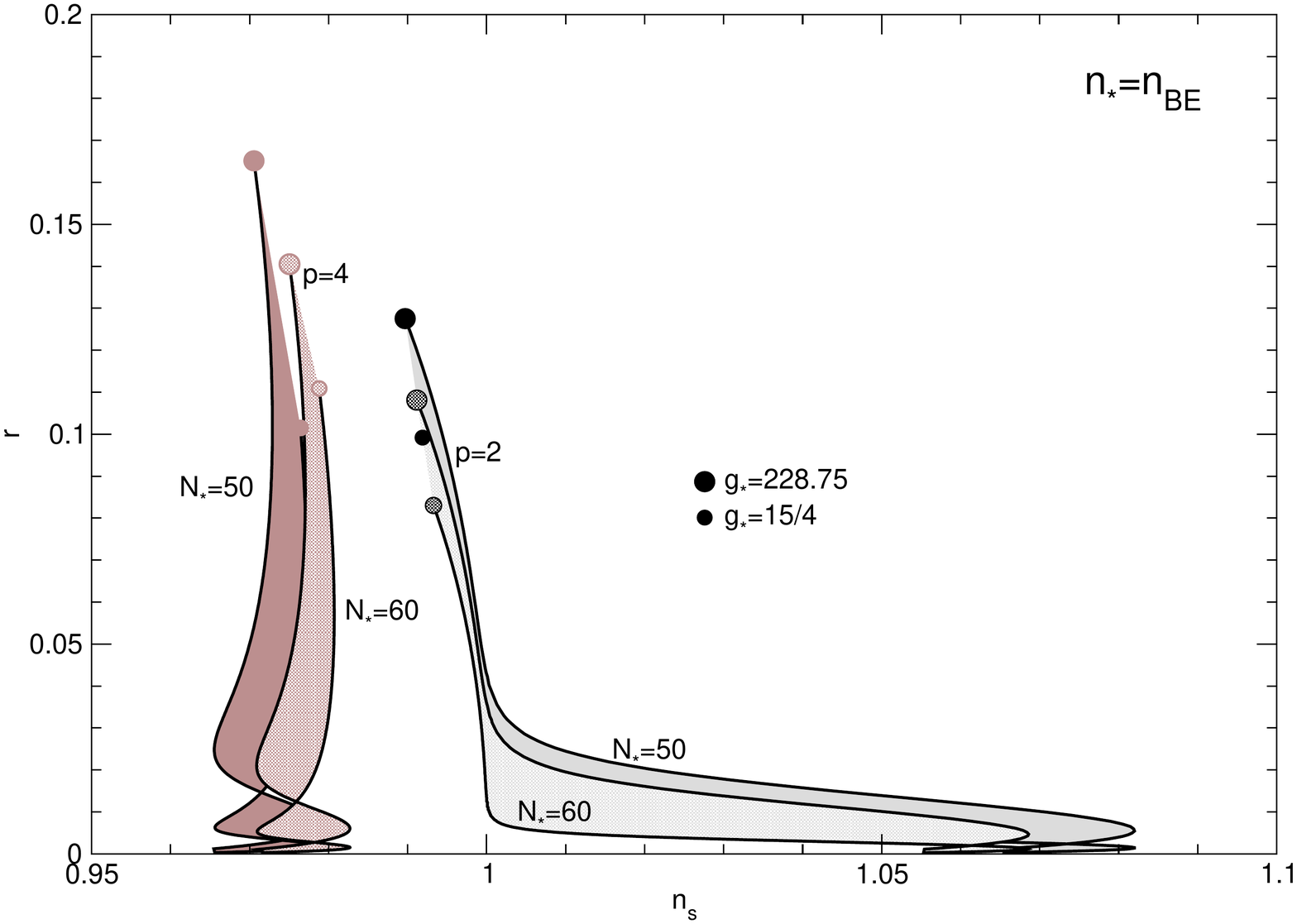} 
\end{tabular}
\caption{Trajectories in the $n_s-r$ plane for the quadratic ($p=2$) and the quartic model ($p=4$), for two choices of the number of e-folds of inflation, $N_*=50,\,60$, as indicated in figure, and $Q_*\lesssim 0.01$. The shaded areas are obtained by varying $g_*$ in the range $[15/4,~228.75]$. The predictions for $T_*=H_*$ are indicated by a circle in each curve. The LHS plot gives the predictions for $n_*=0$., while the RHS plot corresponds to thermalized inflaton fluctuations, $n_*\simeq n_{BE*}(a_*H_*, T_*)$.}
\label{fig1}
\end{figure}

For chaotic models for which $\xi \simeq O(\epsilon,\,\eta)$, the running is always negligible, $|n_s'|\lesssim 10^{-3}$. The spectral index and the
tensor-to-scalar ratio for the quadratic ($p=2$) and the quartic model ($p=4$) are shown in Fig.~\ref{fig1}, with zero inflaton occupation
numbers, $n_*=0$ (LHS plot), and thermal occupation numbers, $n_*\simeq n_{BE}$ (RHS plot). Predictions are plotted for two values of the number of e-folds after horizon crossing, $N_*=50,\,60$. The shaded areas for each curve correspond to varying the number of relativistic degrees of freedom $g_*$ between the minimum $15/4$ ($N_Y=1$) and that of the MSSM, $g_*=228.75$. All the curves start at the value $T_*/H_* \simeq 1$, which gives approximately $Q_*\simeq 10^{-7}$, with both this ratio and $T_*/H_*$ increasing as one moves along the curves up to $Q_*\lesssim 0.01$. 

In the scenario with $n_*=0$, typically the prediction for the tensor-to-scalar ratio is too large, in fact larger than the BICEP2 signal. In this case one has:
\be
r \simeq \frac{16 \epsilon_*}{1 + 2 \pi Q_* T_*/H_*} \,,
\ee
and for $Q_* \lesssim 0.01$ the suppression is not enough to lower the value of $r$. When comparing with cold inflation predictions, the
dominant effect is the increase in the value of $\epsilon_*$, since in increasing the duration of inflation dissipation allows for steeper potentials at horizon-crossing to obtain the desired 50-60 e-folds. Only the case of a quadratic potential, with a small value of $g_*$ and $T_*/H_*\simeq 1$, would be close to the BICEP2 value. 

On the other hand, for thermalized inflaton fluctuations the predictions for the quartic model fall well within the Planck contours, and for $T_*\sim H_*$ the tensor-to-scalar ratio is close to the value measured by BICEP2. The quadratic potential, however, tends to predict a too large spectral index in this case, even for very weak dissipation at horizon-crossing. In this case one has $\epsilon_* \simeq \eta_*\simeq \sigma_*$, making the spectral index blue tilted for most of the parameter space, which is disfavored by the Planck data.

Therefore, the scenario that is most consistent with the current Planck and BICEP2 results is the quartic model with nearly-thermal inflaton occupation numbers, and marginally the quadratic potential for $n_*\simeq 0$ and a small temperature of the thermal bath at horizon-crossing. 

 
\section{Isocurvature modes and effective tensor-to-scalar ratio}

As we have discussed earlier, dissipative effects during inflation result in a net particle production of light degrees of freedom. Whether these correspond to all or only a subset of the (MS)SM degrees of freedom or include light species in additional hidden/sequestered sectors becomes a model-dependent question. From the point of view of the inflationary dynamics, this mainly modifies the parameter $g_*$ determining the relative abundance in radiation, upon which observables exhibit only a mild dependence as illustrated in Fig.~\ref{fig1}.

When dissipation results in the production of the MSSM degrees of freedom, particularly (s)quarks and (s)leptons, an interesting possibility arises. Since these fields result from the decay of the heavy $X$ fields coupled to the inflaton, it is possible that these decays do not conserve baryon number nor the $C$ or $CP$ asymmetries. Baryon number will not be conserved, in particular, if there are at least two decay channels for each $X$ boson or fermion with a different $B$-charge in the final state, while $C$ and $CP$ violation would correspond to complex Yukawa couplings $h$ in the superpotential (\ref{superpotential}) and at least three families of light species. Since dissipation is naturally an out-of-equilibrium process that results in a net particle production, all three Sakharov conditions for baryogenesis can be satisfied. This then yields a possible mechanism for generating a baryon asymmetry during inflation, originally proposed in \cite{BasteroGil:2011cx} and known as warm baryogenesis. 

To realize this mechanism one simply needs to specify the structure of the $Y$ sector and its Yukawa couplings to the $X$ sector in the generic superpotential Eq.~(\ref{superpotential}). A baryon asymmetry results in this case from the decay of heavy bosons or fermions, which is analogous to what is typically considered in other baryogenesis models in the context of GUTs, although the dynamics is inherently different. We may thus consider a set of interactions similar to those found in GUT extensions of the MSSM, such as $SU(5)$ grand unification, yielding a superpotential of the form:
\begin{eqnarray} \label{superpotential2}
W=\sum_{a,i,j}\left[g_a\Phi X_a^2+h_a^{ij}X_aQ_iQ_j+\lambda_a^{ij}X_aQ_i^c L_j\right]~,
\end{eqnarray}
where $Q_i$ and $L_i$ denote in general chiral superfields with a distinct baryonic charge, as for example quark and lepton multiplets as the notation suggests. Note that these Yukawa terms simply realize the generic structure $hXY^2$ for multiple species with distinct Yukawa couplings and carrying different baryon number. The Yukawa matrices $h_a^{ij}$ and $\lambda_a^{ij}$ are complex and will involve non-trivial $CP$ phases if there are at least 3 flavors of the $Q_i$ and $L_i$ multiplets as mentioned above, with $i,j$ denoting the flavor indices. Analogously, one could also realize leptogenesis scenarios of this form, with the $X$ fields corresponding to right-handed (s)neutrinos decaying into (s)leptons and Higgs(ino) fields, thus generating a lepton asymmetry during inflation which could later be converted into a net baryon number by electroweak sphaleron processes.

To compute the net baryon/lepton asymmetry produced during inflation one needs to determine the rate at which particles and anti-particles are produced. Hence, instead of determining the inflaton self-energy as in the computation of the full dissipation coefficient, one can study the self-energy of the individual light particles, as first done in \cite{Graham:2008vu}. The sum of the contributions of all the light species then yields the dissipation coefficient in Eq.~(\ref{dissipation_coefficient}), while the difference between the self-energies of particles and anti-particles gives the asymmetry in their production rates. We refer the reader to the original proposal in  \cite{BasteroGil:2011cx} for the details of this calculation and here we present only the main results relevant for our subsequent analysis. In particular, in the simplest case where the Yukawa couplings for the different flavors only differ by a phase, one obtains the following baryon-to-entropy ratio from bosonic interactions in the slow-roll regime:
\begin{equation} \label{baryon_asymmetry_estimate}
\eta_s={n_b\over s}\approx 8.9\times10^{-11}|h|^2\bigg({T/m_X\over 0.01}\bigg)^2\bigg({\Delta m_X/m_X\over 0.015}\bigg)\bigg({\sin\delta\over0.025}\bigg)~,
\end{equation}
where $\delta$ denotes the effective overall $CP$ phase. One immediately sees that this could account for the observed cosmological asymmetry $7.2\times10^{-11}<\eta_s<9.2\times10^{-11}$ \cite{Fields:2006ga} in the low-temperature regime $T\ll m_X$ with not too small $CP$ phases and $\mathcal{O}(1)$ couplings. We note that, as for GUT baryogenesis models, at least two of the heavy bosons must be non-degenerate, $\Delta m_X\neq 0$, in order to yield non-vanishing interference effects.

Most interestingly, the fact that the final baryon asymmetry depends on the ratio $T^2/m_X^2$ illustrates two attractive features of this model. On the one hand, this dependence arises from the fact that the $X$ bosons are produced off-shell from the inflaton condensate, with the ambient temperature below their mass. On the other hand, in thermal GUT models of baryogenesis and analogous scenarios, an initial population of heavy bosons at $T\gg m_X$ is required, which means that the GUT symmetry must be restored at some point in the cosmological history. By producing the baryon asymmetry through dissipative effects, one thus avoids this requirement and the associated production of heavy monopoles that could come to overclose the universe. 

Moreover, since $T/m_X\propto T/\phi$, the baryon-to-entropy ratio will be slightly different in distinct patches of the universe due to fluctuations of the inflaton field and temperature of the thermal bath, which are related through the slow-roll equations. Although baryons and other light degrees of freedom are subdominant during inflation, they will become a significant component of the universe after matter-radiation equality and thus before the CMB is emitted. This means that super-horizon fluctuations of the inflaton field will also be imprinted in the CMB temperature anisotropies in the form of baryon isocurvature perturbations. These have the same origin and will thus be fully (anti-)correlated with the main adiabatic curvature perturbations. This is a very distinctive feature of warm baryogenesis and makes it a testable model, which is not the case of most of the baryogenesis mechanisms proposed in the literature.

Baryon isocurvature perturbations are characterized by the gauge invariant quantity:
\begin{equation} \label{baryon_iso_def}
S_b={\delta\rho_b\over \rho_b}-{3\over4}{\delta\rho_\gamma\over \rho_\gamma}= {\delta\eta_s\over \eta_s}~,
\end{equation}
and one conventionally considers the ratio of the isocurvature and curvature perturbations on uniform energy density hypersurfaces, $\zeta=-H\delta\rho/\dot\rho$. Using the slow-roll equations one obtains for weak dissipation at horizon-crossing, $Q_*\ll 1$  \cite{BasteroGil:2011cx}:
\begin{eqnarray} \label{baryon_iso_weak}
B_b={S_b\over {\cal \zeta}}= 4\eta_*-6\epsilon_*-6\sigma_*~.
\end{eqnarray}
Note that $B_b>0$ ($<0$) corresponds to fully correlated (anti-correlated) baryon isocurvature modes. One may also determine the spectral index associated with baryon isocurvature modes, obtaining for $Q_*\ll 1$ \cite{Bartrum:2013oka}:
\begin{eqnarray} \label{baryon_iso_index}
n_{iso}=1+{d\log S_b\over d\log k}\simeq n_s + {4\over B_b}\left[2\epsilon_*(2\eta_*-\xi_*)-6\epsilon_*(2\epsilon_*-\eta_*)-3\sigma_*(\sigma_*+2\epsilon_*-\eta_*)\right]~.
\end{eqnarray}
Note that this differs from the scalar (adiabatic) spectral index $n_s$ only by factors of the order of the slow-roll parameters, which may be neglected to leading order so that $n_{iso}\approx n_s$. Since $B_b$ is also given by a combination of the slow-roll parameters, we expect it to be in general of the same order of deviations from scale invariance and the inverse of the number of e-folds of inflation, i.e.~$|B_b|\sim 10^{-2}-10^{-1}$. For example, for the quartic model with thermalized inflaton fluctuations, one obtains $B_b=3(n_s-1)$, yielding $-0.12<B_b<-0.09$ for $0.96<n_s<0.97$. 

One must recall, however, that baryons are not the dominant matter component in the universe and, in particular, the total isocurvature perturbation corresponds to a weighted sum of the contributions from baryons and cold dark matter:
\begin{eqnarray} \label{isocurvature_decomposition}
S_m= {\Omega_c\over \Omega_m} S_c+ {\Omega_b\over \Omega_m} S_b~,
\end{eqnarray}
with $\Omega_c=0.845\Omega_m$ and $\Omega_b=0.155\Omega_m$ according to the best fit Planck model \cite{Ade:2013uln}. Planck has in fact placed bounds on the amount of matter isocurvature perturbations for different types of correlation with the main adiabatic component. For anti-correlated isocurvature perturbations, the bound obtained by Planck for the effective CDM matter isocurvature component corresponds to $|B_b|<0.51$ (95\% CL) in the absence of CDM perturbations  \cite{Ade:2013uln}, so that the quartic potential predictions are well within these bounds. Interestingly, the Planck collaboration has reported a significant improvement of their fit to the CMB temperature data when anti-correlated matter isocurvature modes are included, although no unambiguous evidence for their existence, such as a shift in the position of the acoustic peaks, could be claimed. This is also supported by the more recent analysis in \cite{Kawasaki:2014fwa}, so that the observational effects of matter isocurvature perturbations deserve a closer look.

While our original proposal focused on the production of a baryonic asymmetry, for which there is robust observational evidence, it is clear that the generic model above could lead to the asymmetric production of any light particles carrying a global charge during inflation. In particular, one could easily devise models where the light multiplets include not only the MSSM quarks and leptons but also other chiral multiplets in hidden/sequestered sectors that could account for the observed dark matter abundance in the universe. If such particles carry a global charge that is violated in the decay of at least some of the heavy $X$ fields, a net dark matter asymmetry would be produced. This would give a full {\it warm mattergenesis} mechanism, where both baryons and CDM are produced in an asymmetric fashion during inflation. The asymmetry in the dark matter sector would then be given by an expression of the form in Eq.~(\ref{baryon_asymmetry_estimate}), its value depending on the size of couplings and CP phases in the hidden/sequestered sector. Although no definite predictions can be made, as happens for most if not all mattergenesis models, one can nevertheless say that the baryon and CDM asymmetries will have the same magnitude unless there are large hierarchies between the relevant couplings and phases in the two sectors. This would then explain why the present relative abundance of both components is of the same magnitude provided that dark matter particles have a mass roughly in the few GeV range.

Independently of the value of the potential CDM asymmetry, it is clear that a warm mattergenesis scenario could potentially lead to isocurvature fluctuations in both the baryons and CDM sectors with the same spectrum, since they can be simultaneously produced by dissipation of the inflaton's energy. In the remainder of our analysis we will then consider the possibility of having isocurvature fluctuations in either of these sectors separately or a combined contribution.

CMB temperature anisotropies will in general receive contributions from the main adiabatic component and isocurvature perturbations, as well as primordial gravity waves. Following \cite{Kawasaki:2014lqa, Contaldi:2014zua}, we have: 
\begin{eqnarray} \label{temperature_variance}
\langle (\Delta T/T)^2\rangle\sim P_{\zeta} + 4P_{S_m}+4 P_{{\cal \zeta} S_m}+{5\over6 }P_t~,
\end{eqnarray}
where $P_i$ refers to the power spectrum of each component and $P_{{\cal \zeta} S_m}$ to the correlation between scalar adiabatic and isocurvature modes.  We may then express this as:
\begin{eqnarray} \label{temperature_variance_2}
\langle (\Delta T/T)^2\rangle\sim P_{\zeta} \left(1+4 B_m^2 +4B_m+{5\over6}r\right)~,
\end{eqnarray}
where, as in the pure baryonic case, the sign of $B_m$ determines whether isocurvature and adiabatic modes are fully correlated or anti-correlated.  In  \cite{Kawasaki:2014lqa}, it was shown that in the anti-correlated case, isocurvature perturbations can completely cancel the contribution from tensor modes to the CMB temperature power spectrum for $B_m\simeq 0.04$. This is in agreement with the Planck bound ($|B_m|<0.079$) and the earlier less stringent WMAP3 bound  $|B_m|<0.13$ \cite{Kawasaki:2007mb}) for anti-correlated modes. This would then be able to accommodate a large tensor-to-scalar ratio as suggested by the BICEP2 detection, to which the temperature power spectrum would be completely insensitive.

Here we pursue a less drastic approach, where we note that isocurvature modes may only partially screen the tensor contribution. If the temperature power spectrum is analyzed assuming only the presence of adiabatic modes and a (subdominant) tensor component, one can only determine an effective tensor-to-scalar ratio: 
\begin{eqnarray} \label{temperature_variance_eff}
\langle (\Delta T/T)^2\rangle\sim P_{\zeta} \left(1+{5\over6}r_\mathrm{eff}\right)~,
\end{eqnarray}
where
\begin{eqnarray} \label{r_eff}
r_\mathrm{eff}=r +{6\over5}\left(4B_m^2+4B_m\right)
\end{eqnarray}
includes the contribution from isocurvature modes, which is not accounted for explicitly in an adiabatic+tensor model of the data. It is then easy to see that for anti-correlated modes $r_\mathrm{eff}<r$, so that it may be consistent to have $r_\mathrm{eff}<0.11$, while $r>0.11$ and close to the value inferred from the BICEP2 data. The complete screening considered in \cite{Kawasaki:2014lqa} corresponds to $r_\mathrm{eff}=0$, but the only true observational constraint is the Planck bound $r_\mathrm{eff}<0.11$  \cite{Ade:2013uln} . 

We may then use Eq.~(\ref{baryon_iso_weak}) in Eq.~(\ref{r_eff}) to determine the effective tensor-to-scalar ratio that would be inferred from the CMB temperature spectrum for different inflationary models. Given the predictions in the $n_s-r$ plane obtained in the previous section, we will mainly focus on the quartic model with thermal inflaton phase-space distribution, which as discussed in \cite{Bartrum:2013fia} yields a remarkable agreement with the Planck results. We will nevertheless show predictions including isocurvature modes for the quadratic model in both the negligible and nearly-thermal inflaton occupation numbers.

\begin{figure}[htbp]
\centering\includegraphics[scale=1]{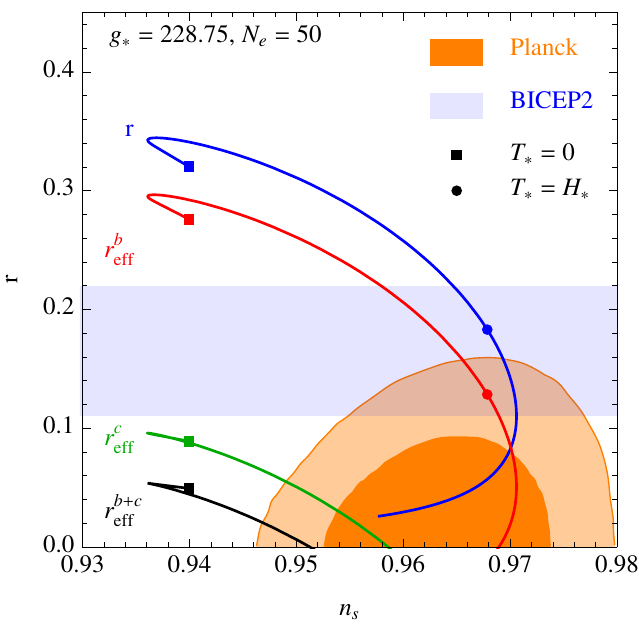}\hspace{1cm}
\centering\includegraphics[scale=1]{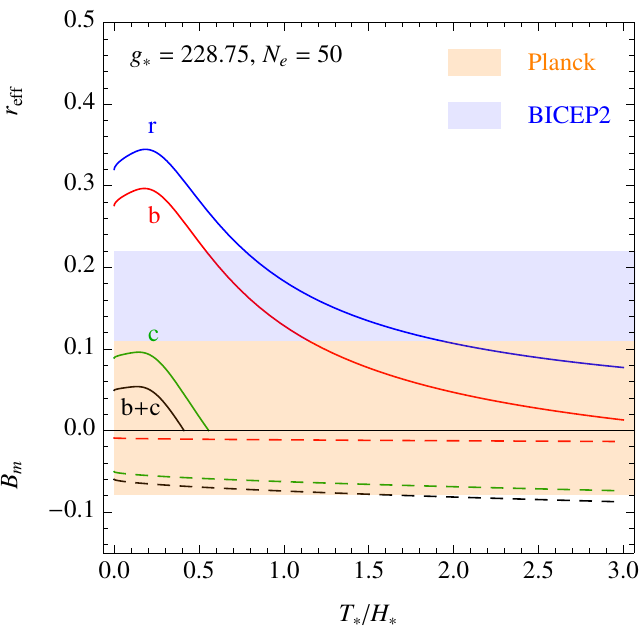}
\caption{LHS: Trajectories in the $n_s - r$ plane for the quartic model (thermal inflaton occupation numbers) with $50$ e-folds of inflation and $g_*=228.75$, with baryon ($r_{eff}^b$), CDM ($r_{eff}^c$), and full matter isocurvature perturbations ($r_{eff}^{b+c}$). The shaded regions show the 68\% and 95\% CL Planck contours \cite{Ade:2013uln}, including the results of WMAP and BAO observations, and the 1$\sigma$ interval for $r$ derived by the BICEP2 collaboration \cite{Ade:2014xna} after taking into account dust contributions. These results correspond to the range $T_*/H_*=0 - 13$ ($Q_*=0-0.01$). RHS: Associated predictions for $r$, $r_\mathrm{eff}^i$  (upper solid curves) and $B_m^i$, $i=b,\,c,\,c+b$ (lower dashed curves), as a function of $T_*/H_*$. Shaded areas give the 1$\sigma$ regions for $r$ from BICEP2 and the $95\%$ CL limits from Planck on $r_\mathrm{eff}$ (positive band) and anti-correlated isocurvature modes (negative band).}
\label{figpp4thiso}
\end{figure}

For the quartic model, $V(\phi)=\lambda\phi^4$, with nearly-thermal fluctuations, the trajectories in the $n_s-r$ plane as a function of $T_*/H_*$ are illustrated in Fig.~\ref{figpp4thiso} (left plot),  for 50 e-folds of inflationary expansion after horizon-crossing of the Planck/BICEP pivot scale and $g_*=228.75$  (all MSSM degrees of freedom in the thermal bath), which yields the best agreement with the data. In this figure we show the physical tensor-to-scalar ratio and the effective value corresponding to screening by baryon, CDM or full matter isocurvature perturbations (we refer the reader to the analysis in \cite{Bartrum:2013fia} and references therein for more details on the dynamics of warm inflation in this model). 

Note that in this figure we have used the expressions for the observables valid for arbitrary $T_*/H_*$, although one should take into account that for $T_*<H_*$ there may be significant de-Sitter corrections not only to the inflaton fluctuation distribution but also to the dissipation coefficient, as discussed earlier. Nevertheless, the curves illustrate the connection between the zero-temperature predictions (which are incompatible with the Planck results) and the warm inflation results, for $T_*\gtrsim H_*$, which may be in agreement with both Planck and BICEP. Note that this requires $n_s$ and $r_\mathrm{eff}^{i}$, for $i=b,\,c,\, b+c$  (red, green, black curves) to be within the Planck contours, while $r$ (blue curve) must be within the BICEP2 window.  In this case, baryon isocurvature modes are sufficient to screen the tensor-to-scalar ratio by an amount $\Delta r_\mathrm{eff}=r-r_\mathrm{eff}=0.04-0.06$, which as one can see is enough to reconcile the BICEP2 observations with the Planck upper bound $r<0.11$ (95\% CL). On the other hand, CDM isocurvature modes already give a too large correction to the spectrum, rendering the corresponding $r_{eff}$ indeed negative, being thus disfavored. 

To better illustrate this result, in Fig.~\ref{figpp4thiso} (right plot) we also show the variation of $r$, $r_\mathrm{eff}^i$ and $B_m$ with the value of $T_*/H_*$. We also include in this figure the corresponding BICEP2 $1\sigma$ window and the $95\%$ CL limits placed by Planck on the effective tensor-to-scalar ratio and matter isocurvature perturbations. 

This figure clearly shows that for $T_*/H_*\simeq 1.2-2$ the effective tensor-to-scalar ratio screened by isocurvature modes is within Planck's upper bound, while the true value of $r$ is compatible with the BICEP2 result. For a larger number of e-folds, the tension between the two experiments is removed for smaller values of $T_*/H_*$, with for example $N_e=60$ yielding $T_*/H_*=1-1.6$. For $N_e=40-60$, agreement between Planck and BICEP2 for the tensor-to-scalar ratio yields $n_s=0.962-0.975$, also within the Planck bounds. While the contribution of anti-correlated baryon isocurvature modes is comfortably within the Planck window, $|B_m|<0.079$, CDM yields a much larger contribution that is very close to Planck's upper limit, which could be ruled out in the near future. As mentioned above, the corresponding effective tensor-to-scalar ratio is negative for $T_*\gtrsim H_*$, which for clarity we have excluded from Fig.~\ref{figpp4thiso}.

\begin{figure}[!t]
\begin{tabular}{ccc}
\centering\includegraphics[scale=1]{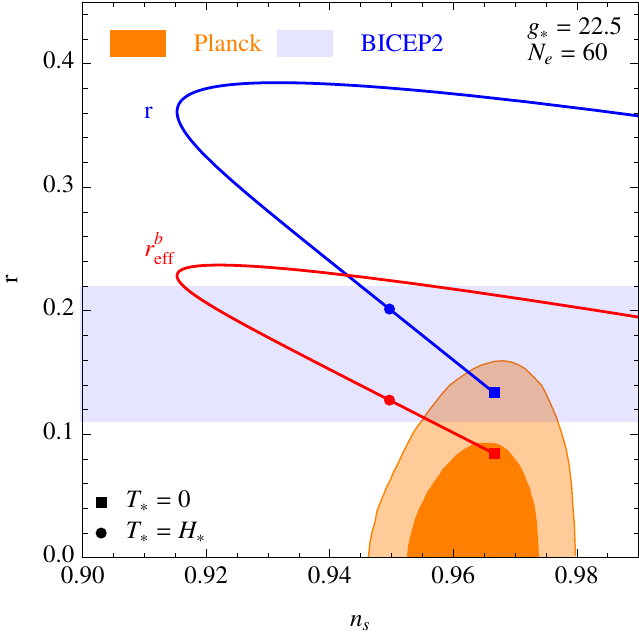} && \hspace{1cm}
\centering\includegraphics[scale=1]{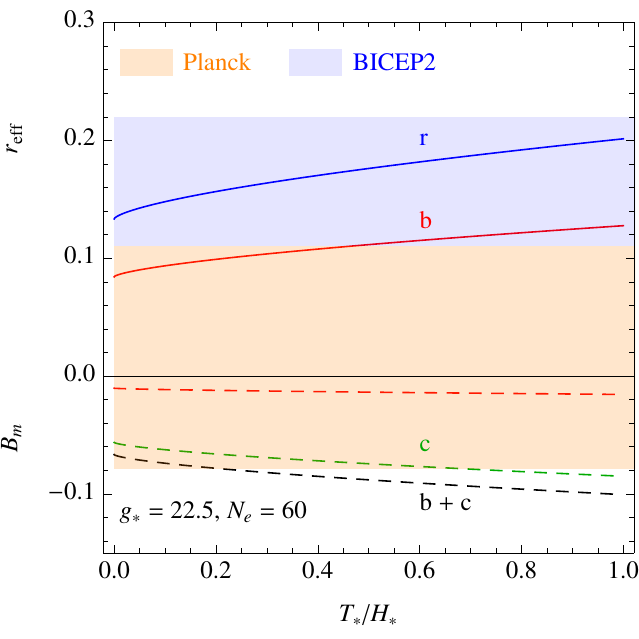} 
\end{tabular}
\caption{LHS: Trajectories in the $n_s - r$ plane for the quadratic model (negligible inflaton occupation numbers), for $60$ e-folds of inflation and $g_*=22.5$, with baryon isocurvature perturbations $r_{eff}^b$.  The same conventions as in Fig.~\ref{figpp4thiso} are used. RHS: Associated predictions for $r$, $r_\mathrm{eff}^b$  (upper solid curves) and $B_m^i$, $i=b,\,c,\,c+b$ (lower dashed curves), as a function of $T_*/H_*$. Shaded areas are the 1$\sigma$ regions for $r$ from BICEP2 and the $95\%$ CL Planck upper limits (bottom).} 
\label{figpp2nonthiso}
\end{figure}

In Fig.~\ref{figpp2nonthiso} we have done a similar analysis for the quadratic chaotic model, $V(\phi)=m^2\phi^2/2$, considering the opposite limit of negligible inflaton occupation numbers, where the spectrum is modified essentially by the FD contribution. The results in Section II suggested that a larger number of e-folds and a lower number of relativistic species would yield a better agreement with the data, in particular a lower tensor-to-scalar ratio, so that we take $N_*=60$ and $g_*=22.5$, the latter corresponding to three families of quark and lepton chiral multiplets, which is the minimum required for the generation of a baryon asymmetry.

Matter isocurvature modes are also anti-correlated with the main adiabatic component for this potential, as is in fact the case for all monomial potentials with power $p<10$. For the quadratic model, given that the FD correction tends to increase the value of $r$ for not too large values of $T_*/H_*$, baryons alone are not able to bring down $r_{eff}$ by a sufficiently large amount. On the other hand, the larger contribution from CDM, or the combination of both baryons and CDM, completely screens the tensor modes and renders  $r_{eff}$ negative, so that these curves are not shown in Fig.~\ref{figpp2nonthiso}. Moreover, the CDM contribution to $B_m$ falls outside Planck's upper bound for $T_*>0.6H_*$. One may notice that the $T_*=0$ predictions including baryon isocurvature modes fall within the 1$\sigma$ contours. However, in this limit there is no dissipation and a baryon asymmetry cannot be produced. Nevertheless, it may be possible to have instead dissipation with $T_*/H_* \lesssim 1$, which may render $r_{eff}^b$ consistent with the present Planck value at 1$\sigma$-2$\sigma$ level (red line from the square to the circle in Fig.~\ref{figpp2nonthiso}).  This case requires, however, further exploration since as mentioned above no analysis of dissipative effects in this limit has been performed in the literature to our knowledge. We note that a similar behavior is observed for the quartic model with negligible inflaton occupation numbers in the thermal bath, but in this case the values of $r$ are well above the allowed BICEP2 window, as can be seen in Fig.~\ref{fig1} (left plot).

\begin{figure}[htbp]
\begin{tabular}{ccc}
\centering\includegraphics[scale=1]{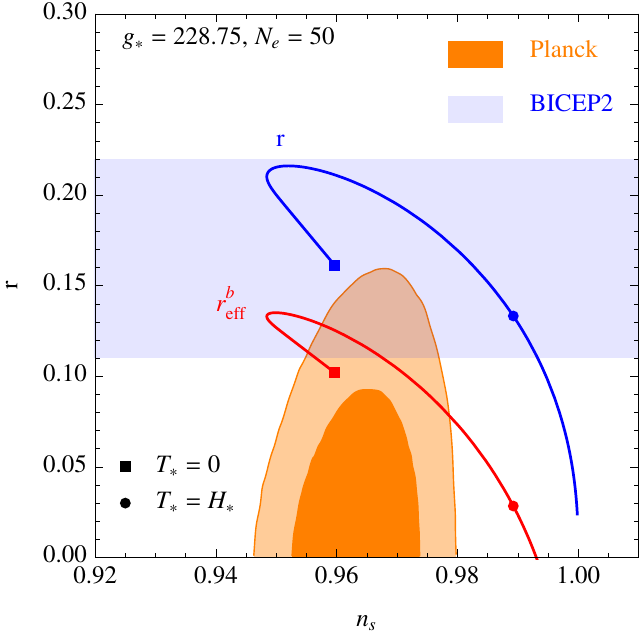} \hspace{1cm}
\centering\includegraphics[scale=1]{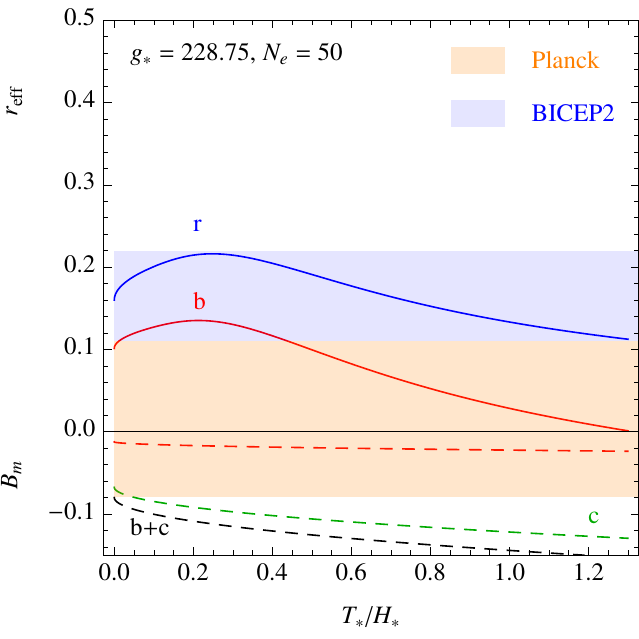} 
\end{tabular}
\caption{LHS: Trajectories in the $n_s - r$ plane for the quadratic model (thermal inflaton occupation numbers), for $50$ e-folds of inflation and $g_*=228.75$, with baryon isocurvature perturbations $r_{eff}^b$.  The same conventions as in Fig.~\ref{figpp4thiso} are used. RHS: Associated predictions for $r$, $r_\mathrm{eff}^b$  (upper solid curves) and $B_m^i$, $i=b,\,c,\,c+b$ (lower dashed curves), as a function of $T_*/H_*$. Shaded areas are the 1$\sigma$ regions for $r$ from BICEP2 and the $95\%$ CL Planck upper limits (bottom).} 
\label{figpp2thiso}
\end{figure}

A similar effect in the region $T_*\lesssim H_*$ is observed for the quadratic model in the opposite limit of thermal inflaton fluctuations, as can be seen in Fig.~\ref{figpp2thiso}. Again, although dissipation with $T_*/H_* \geq 1$ pushes the model towards too large values of the spectral index,
outside the present observational window, small thermal effects might affect the observables in a non-trivial way even for $T_*\lesssim H_*$,
bringing the prediction of the spectral index closer to that of cold inflation. If this also allows for the generation of a baryon asymmetry through dissipation, the associated isocurvature perturbation will also play a role when constraining the value of the tensor-to-scalar ratio.

This analysis shows that the quartic model with nearly-thermalized inflaton fluctuations yields the best agreement with the data, with the baryon isocurvature modes associated with warm baryogenesis screening the tensor-to-scalar ratio sufficiently to reconcile the present BICEP2 detection with last year's Planck results. 

CDM isocurvature modes in the context of a more general mattergenesis mechanism generically yield a too large screening that would reduce the overall power ($r_\mathrm{eff}$) and are hence in more tension with the present data. However, this assumes that all dark matter in the universe is charged under a global symmetry similarly to the baryonic sector. If, on the contrary, the dark sector is composed of multiple species and only a sub-sector is produced asymmetrically during inflation, the CDM contribution to matter isocurvature perturbations will be smaller than computed above. The corresponding screening of the tensor-to-scalar ratio will thus be parametrically reduced and yield a better agreement with the data for the scenarios outlined above, as e.g.~for the quadratic potential with negligible inflaton occupation numbers. Hence, depending on the fraction of CDM produced asymmetrically during inflation, the effective tensor-to-scalar ratio may vary from the true value, when no baryon or CDM asymmetry is produced during inflation, to the fully screened value when an asymmetry is produced for the total baryon and dark matter abundance.

\begin{figure}[htbp]
\begin{tabular}{ccc}
\centering\includegraphics[scale=1]{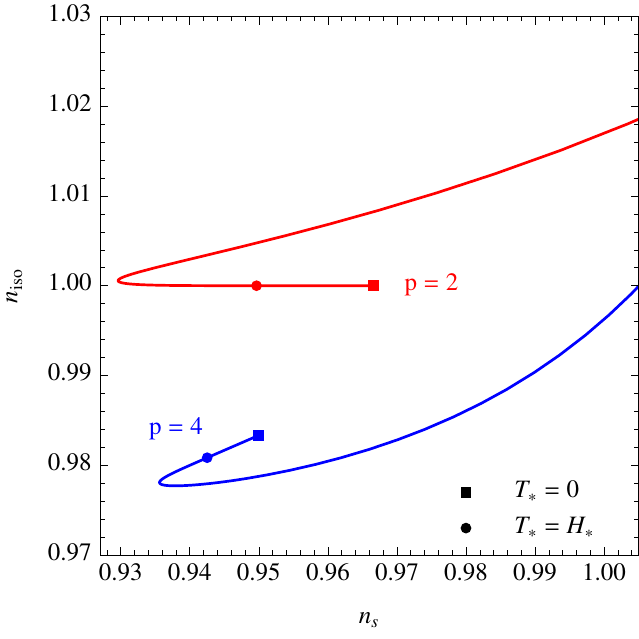} \hspace{1cm} &&
\centering\includegraphics[scale=1]{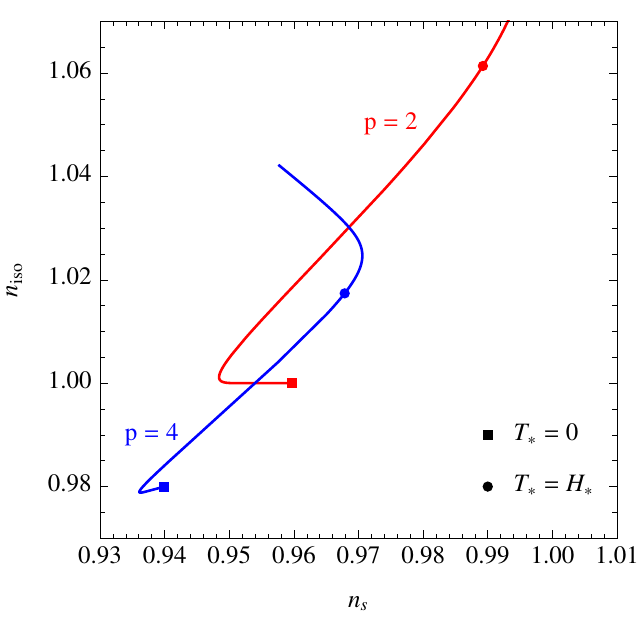} 
\end{tabular}
\caption{Spectral indices for adiabatic and isocurvature perturbations as a function of $T_*/H_*$ for the quartic ($p=4$) and quadratic ($p=2$) potentials, considering (LHS) negligible inflaton occupation numbers  ($g_*=22.5$, $N_e=60$) and (RHS) thermal inflaton occupation numbers ($g_*=228.75$, $N_e=50$).} 
\label{figniso}
\end{figure}

Finally, we note that the analysis performed by the Planck collaboration on the presence of anti-correlated isocurvature modes assumed $n_{iso}=n_s$, which is only approximately true for warm baryogenesis/mattergenesis.  In Fig.~\ref{figniso} we illustrate the difference between the two spectral indices for chaotic models in the two limits of negligible and nearly-thermal inflaton occupation numbers with the observationally preferred parameters in each case.

As one can see, $n_{iso}$ tends to be slightly larger than $n_s$ in these models, with differences of $\mathcal{O}(10^{-2})$, which may not be easy to distinguish observationally. The isocurvature spectral index could nevertheless provide an additional observable to test the warm baryogenesis/mattergenesis scenario and more generically the warm inflation paradigm, yielding for example additional consistency relations along the lines proposed in \cite{Bartrum:2013oka}.
%

\section{Conclusion}

There have been several proposals in the literature to explain the BICEP-Planck discrepancy, essentially by including other parameters in
the model describing the temperature power spectrum. The BICEP2 collaboration was in fact the first to propose the inclusion of a
significant running of the scalar spectral index \cite{Ade:2014xna}, which seems however difficult to find within conventional inflationary models. Other interesting proposals include for example novel light species such as sterile neutrinos \cite{Zhang:2014dxk}. However, the inclusion of isocurvature modes in the context of warm baryogenesis or mattergenesis is, in our opinion, particularly attractive, since it may provide a unique observational window into the generation of baryonic and/or cold dark matter with very distinct observational signatures.

Our results for chaotic models suggest that a large tensor-to-scalar ratio could be accommodated by the current Planck results due to the presence of matter isocurvature modes associated with an asymmetric dissipation of the inflaton's energy density into baryonic or CDM species. Typically the latter give a too large screening of the tensor-to-scalar ratio in the temperature power spectrum and are already in tension with the current Planck bounds in a few cases, with baryonic perturbations yielding the most promising avenue. Nevertheless, CDM isocurvature modes may yield a parametrically smaller screening if an asymmetry is produced in only a sub-sector of the dark matter species. The CMB temperature power spectrum does not distinguish between baryonic and CDM perturbations, but it is possible that 21 cm line observations could break this degeneracy in the near future \cite{Kawasaki:2011ze}, thus helping to determine which type of asymmetries could be produced during inflation.

The analysis performed in this work shows that isocurvature modes from warm baryogenesis/mattergenesis could yield a significant effective contribution to the spectrum of CMB temperature anisotropies. Naturally, the next step should be to include these components explicitly in the analysis of the temperature data. A study along these lines has recently been performed in \cite{Kawasaki:2014fwa}, combining the BICEP2 and Planck data, and which reinforces the preference for the inclusion of anti-correlated matter isocurvature modes that was already  observed in last year's Planck analysis \cite{Ade:2013uln}. A preliminary comparison shows that the predictions of warm baryogenesis/mattergenesis for the quartic model with nearly-thermal inflaton fluctuations are in good agreement with this analysis, although the authors of \cite{Kawasaki:2014fwa} have yet to consider the effects of astrophysical foreground subtraction in the BICEP2 data. We point out that, to the best of our knowledge, warm baryogenesis/mattergenesis is the only model where a full (anti-)correlation between adiabatic and isocurvature modes is naturally generated.

We are experiencing exciting times for research in inflationary dynamics, with the results of the BICEP2 experiment leading the way to really test the inflationary hypothesis. The BICEP2 results point towards the inclusion of additional ingredients in the simplest single-field inflation models, but the claim for the observation of primordial gravity waves with a large amplitude will remain uncertain until other experiments are able to confirm this result and exclude other possible sources of astrophysical B-mode polarization \cite{Liu:2014mpa}.   

Warm inflation provides a complete change of paradigm with respect to conventional supercooled scenarios, introducing significant modifications to the dynamics of inflation and its observational predictions, as well as the role of the inflaton in addressing other cosmological problems. As in the supercooled framework, observational predictions in warm inflation depend on the particular inflationary potential considered. As we have illustrated in this work, even chaotic models could yield a range of values for the tensor-to-scalar ratio, from very small to above the level claimed by BICEP2, depending essentially on the ambient temperature at horizon-crossing. In fact, if the observed tensor-to-scalar ratio turns out to be
below 0.1, the warm quartic model will be amongst the most compelling. Other potentials such as in SUSY hybrid inflation, where the waterfall fields mediate dissipative effects, typically yield a small tensor-to-scalar ratio, and for example the predictions for the spectral index are in much better agreement with the Planck data than in the corresponding supercooled regime \cite{Bastero-Gil:2013owa}. We would thus like to emphasize that, although particularly relevant to address the present tension between the Planck and BICEP results, baryon and CDM isocurvature modes may have a significant effect independently of the true gravity wave abundance produced during inflation.

Only time and further scrutiny will tell if the tensor-to-scalar ratio is really as large as BICEP2 suggests, but it is clear that individual observables cannot break the degeneracy between supercooled and warm inflation scenarios. However, the existence of additional effects such as isocurvature perturbations or a potentially observable primordial non-gaussianity, as well as consistency relations between these different observables \cite{Bartrum:2013oka}, yield a promising avenue to accurately determine the paradigm that best describes the inflationary universe. 

Moreover, we note that interactions between the inflaton and other fields are always required in a complete inflationary model, since a `graceful exit' into the standard Big Bang evolution must always be attained after inflation. Whether inflation occurs in a supercooled or warm regime depends parametrically on the couplings and field multiplicities involved, which determine whether a radiation bath can be sustained all through inflation or if it can only be (re-)created at the end of the slow-roll evolution. If, on the one hand, warm inflation may seem to require more parameters to describe the slow-roll dynamics and associated observables, these are in fact only relegated to describing a separate reheating period in supercooled scenarios. On the other hand, as we have discussed in this work, this opens up the possibility for new physical processes to occur during inflation, with additional observables and distinctive features that provide us with a unique opportunity to probe the full particle physics description behind the inflationary universe.

The emerging picture after the recent results of Planck and then BICEP2 shows that warm inflation can be consistent with the data for the simplest and most robust monomial potentials, in particular the quartic potential. The warm inflation models analyzed in this article arise from full first principles quantum field theory calculations \cite{BasteroGil:2010pb, BasteroGil:2012cm} and much of what has been applied here was developed over the past several years alongside the cosmological applications to warm inflation.  More recently, studies focused primarily on reheating have emerged, which have relevance to warm inflation, thermalization \cite{Mukaida:2012qn, Mukaida:2012bz, Drewes:2013iaa, Enqvist:2013qba, Harigaya:2013vwa} and dissipation \cite{Mukaida:2012qn, Mukaida:2012bz, Mukaida:2014yia}. Moreover, the same fluctuation-dissipation dynamics relevant for warm inflation has applicability for reheating \cite{BasteroGil:2010pb}, curvaton dynamics and other early universe scenarios \cite{Bastero-Gil:2014jsa}. There is a growing understanding of the dynamics of warm inflation and further developments will eventually allow for an embedding of warm inflation into concrete particle physics models of the early universe. This will ultimately provide a full dynamical description of not only inflation but also other cosmological problems, such as baryogenesis or dark matter,  with overlaps between these areas, as we have demonstrated in this work.

\section*{Acknowledgements}

A.B.~is supported by STFC. M.B.G.~is partially supported by ``Junta de Andaluc\'ia'' (FQM101) and would like to thank the hospitality of the LPSC, Grenoble, during the completion of this work.  R.O.R.~is partially supported by research grants from Conselho Nacional de Desenvolvimento Cient\'{\i}fico e Tecnol\'ogico (CNPq) and Funda\c{c}\~ao Carlos Chagas Filho de Amparo \`a Pesquisa do Estado do Rio de Janeiro (FAPERJ). J.G.R,~is supported by FCT (SFRH/BPD/85969/2012) and partially by the grant PTDC/FIS/116625/2010 and the Marie Curie action NRHEP-295189-FP7-PEOPLE-2011-IRSES. 


\bibliographystyle{model1-num-names}

\end{document}